\documentclass[aps,twocolumn,prl,showpacs,color,psfig,epsf]{revtex4}
\usepackage{graphicx}
\usepackage{amsmath,amssymb,latexsym,color,amsfonts}
\bibstyle{apsrev.bib}

\newcommand{\beq}{\begin{equation}}
\newcommand{\eeq}{\end{equation}}

\begin{document}
\title{Shear banding, aging and noise dynamics in soft glassy materials}
\author{S. M. Fielding,$^1$ M. E. Cates,$^2$ P. Sollich,$^3$}
\affiliation{
$^1$School of Mathematics and Manchester Centre for Nonlinear
Dynamics, University of Manchester, Oxford Road, Manchester M13 9EP, UK \\
$^2$SUPA, School of Physics, University of Edinburgh, JCMB Kings Buildings,
Edinburgh EH9 3JZ, UK \\
$^3$Department of Mathematics, King's College London, Strand, London WC2R 2LS, UK}

\begin{abstract}
The `soft glassy rheology' (SGR) model gives an appealing account of the flow of nonergodic soft materials in terms of the local yield dynamics of mesoscopic elements. Newtonian, power-law, and yield-stress fluid regimes arise on varying a `noise temperature', $x$. Here we extend the model, to capture the idea that the noise is largely caused by yield itself. The extended model can account for the viscosity-bifurcation and shear-banding effects reported recently in a wide range of soft materials. A variant model may shed light on shear banding and strain-rate hysteresis seen in glassy star polymers solutions.
\end{abstract}

\maketitle

Nonergodic materials, such as colloidal gels and dense emulsions, have unusual rheology that is exploited in many commercial applications such as paints, foodstuffs, and cleaning products \cite{coussotreview}. Understanding their flow behavior represents a profound challenge to statistical physics: this is intractable in its full generality, but promising progress has recently been made at three levels. Fully microscopic work -- based for instance on mode-coupling theory (MCT) -- is so far limited mainly to monodisperse colloidal glasses \cite{catesfuchs}; it remains technically formidable. Purely phenomenological approaches, in which a continuum stress or strain representation of the local state is supplemented by just one or two variables (e.g.~`fluidity') \cite{picard,coussot} are much simpler but of rather limited predictive power, although some recent developments create a more precise link between microscopic physics and the structural variables that control the continuum behaviour. (An example is the recent use of a time-evolving non-affine parameter in a model of star-polymer rheology \cite{beris}.)
The middle ground is represented by mesoscopic models \cite{coussotreview,falk,SGR}, in which explicit but coarse-grained degrees of freedom obey some specified dynamics. 

Among mesoscopic approaches, the SGR model \cite{SGR} has become widely used to interpret experiments on soft matter, biophysics, and even conventional glasses \cite{coussotreview,followups}. The qualitative  successes of SGR include the prediction of power-law ($\sigma\sim\dot\gamma^{x-1}$) and Herschel-Bulkley ($\sigma-\sigma_Y\sim\dot\gamma^{1-x}$) regimes for steady-state flow curves, where $\sigma$ is shear stress, $\sigma_Y$ a dynamic yield stress and $\dot\gamma$ strain rate. These regimes are controlled by an effective noise temperature, $x$, which governs the jump rate out of local traps. In the yield stress regime ($x<1$) SGR also predicts rheological aging, primarily via a so called `simple aging' scenario \cite{SGRage,aging}. 
This is something that MCT cannot yet capture \cite{catesfuchs}. On the other hand, an important drawback of SGR is that its mesoscopic physics (trap-hopping) is somewhat generic, with few clues as to how the parameters of the model should be varied to address different classes of physical systems  -- for example, hard-sphere versus soft-sphere interactions. (Purely phenomenological models also suffer from this drawback, whereas MCT takes as input the equilibrium structure factor which can account for such differences.) Another drawback of SGR is that it does not readily admit nonmonotonic flow curves of the type that would lead to shear banding -- the coexistence at fixed $\sigma$ of layers with different $\dot\gamma$, or vice versa \cite{foothead}. Such limitations may be linked to the fact that the SGR model takes $x$ as a constant model parameter, while in practice the noise temperature for a given trap should depend on the level of jump activity in its vicinity. 

Currently we are not in a position to derive the $x$ dynamics from first principles, nor even mesoscopic ones. Rather, in what follows we treat the $x$ evolution at a phenomenological level (rather as fluidity is treated in \cite{picard,coussot}). We will show that even a very simple choice can then account for the `viscosity bifurcation' seen in many soft materials \cite{coussot,bonn}. On applying a step stress, there is a stress threshold below which the system remains solid, but above which it flows homogeneously at a shear rate that exceeds a finite minimum value; we find moreover that the critical stress for the viscosity bifurcation  itself depends on sample history prior to the application of step stress.
At imposed shear rate, the same physics results in coexistence of a rigid, aging, `cold' band ($x=x_1<1$) and a fluid, ergodic, `hot' band ($x=x_2>1$) at a common stress $\sigma$ obeying $0=\sigma_Y(x_2)<\sigma<\sigma_Y(x_1)$ \cite{coussot2}. Within a different variant of our model, we also find below a distinctive hysteretic shear-banding scenario reminiscent of one reported recently in star polymer solutions \cite{stars}. Thus our work points to a possible new connection between glass-based descriptions of nonergodic matter and two major experimental scenarios, complementing previous, purely phenomenological, modelling for these \cite{coussot,stars2}. However, our approach remains semi-phenomenological in that the choice of model variant and/or parameters cannot be linked directly to microstructure.

The SGR model starts from a trap dynamics for mesoscopic elements. The jump rate out of a given trap is $\Gamma_0\exp[-(E-k\ell^2/2)/x]$, with $\Gamma_0$ an attempt rate, $E$ the well depth, $k$ a (uniform) elastic constant, and $\ell$ the local strain. The latter evolves between jumps as $\dot\ell = \dot\gamma$ representing affine shearing by macroscopic strain. New traps have (for simplicity) $\ell = 0$; they have $E$ values drawn from a prior distribution $\rho(E)\propto e^{-E}$ whose form is chosen, following Bouchaud \cite{aging}, to engineer an arrest transition, which occurs at $x=1$. As stated previously $x$ is viewed as an effective noise temperature \cite{SGR}. However a clear physical interpretation of this parameter remains lacking. One idea is that a thermal nonergodic system will, on cooling, hover in a `marginal state' with an effective temperature close to that of its glass transition \cite{marginal}; another is that $x$ is mechanical noise created by plastic rearrangements elsewhere in the system. The latter implies a coupling between $x$ and the flow dynamics, which was acknowledged but neglected in previous work (except \cite{foothead}), and whose consequences we now explore.

To do so, we start with the SGR equation for the trap probability distribution $P(E,\ell,t)$
\begin{equation}
\dot P = -\dot\gamma\frac{\partial P}{\partial \ell} - \Gamma_0
e^{-(E-k\ell^2/2)/x}P + \Gamma\rho(E)\delta(\ell)\label{sgr}
\end{equation}
(with $\Gamma = \Gamma_0 \langle e^{-(E-k\ell^2/2)/x}\rangle_P$ the total jump rate) and couple this to a relaxation-diffusion dynamics for $x$:
\begin{equation}
\tau_x \dot x(y) = - x(y) +x_0 + {\cal S} + \lambda^2\frac{\partial^2 x}{\partial y^2}= 0 \label{xdynamics} \end{equation}
We have assumed for simplicity that $x$ and $\dot\gamma$ depend on a single spatial coordinate $y$ in the shear gradient direction. The source term ${\cal S}(y)$, which represents the pumping of noise by jump events, clearly depends on $P(E,\ell,t)$ at position $y$, which we henceforth denote $P(y)$. 

\begin{figure}
\centerline {\includegraphics[width=2.8in]
{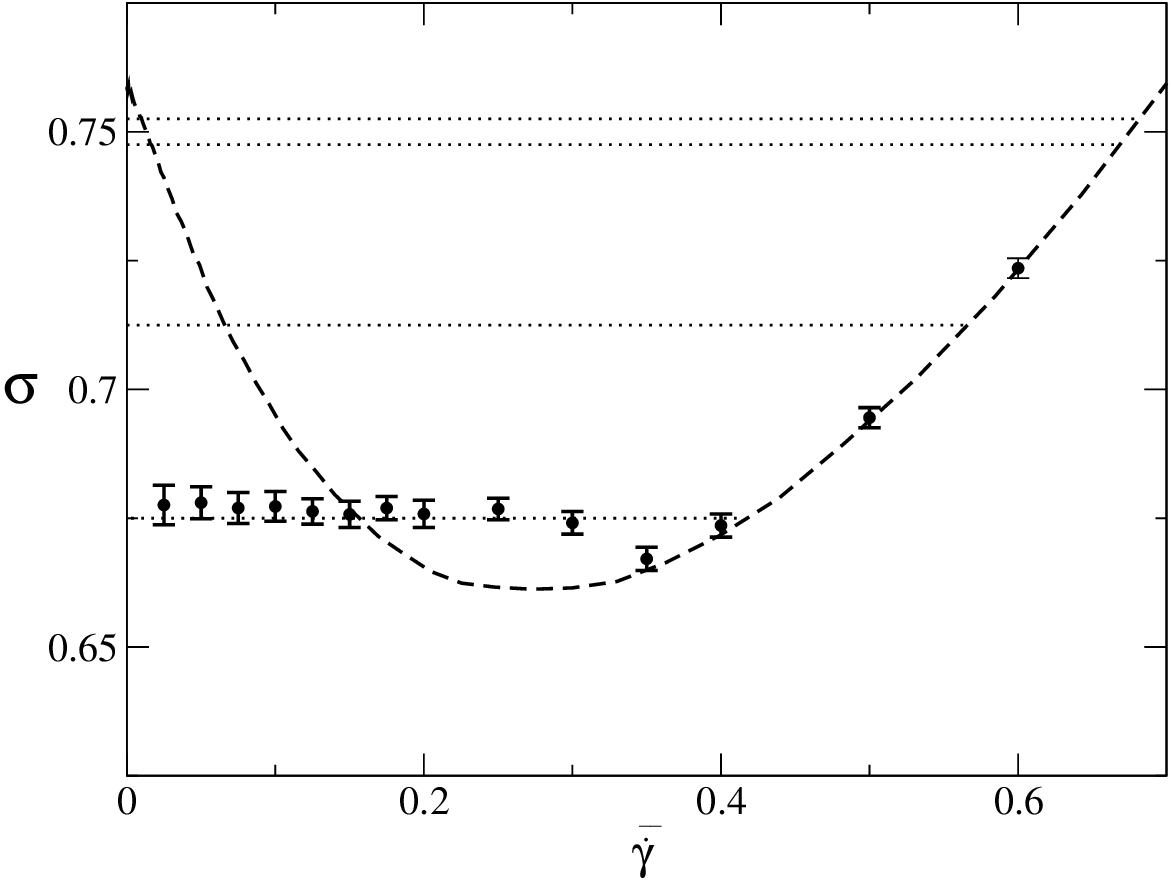}}
\caption{Dashed line: constitutive curve $\sigma(\dot\gamma)$ for $x_0 = 0.3, a= 2.0$. Symbols: WTMC results for quasi-stationary stress $\sigma$ at various imposed mean strain rates $\overline{\dot\gamma}$ found after start up of steady shear. Shear banding is present whenever the WTMC data deviates significantly from the dashed curve. Dotted lines: viscosity bifurcation points ($\pm 0.05$) in step stress for $\log_{10}t_w = 1.0,1.5,2.0,2.5$ (bottom to top); {\em cf} Fig.\ref{fourth}. }
\label{first}
\end{figure}

\begin{figure}
\centerline {\includegraphics[width=2.6in]
{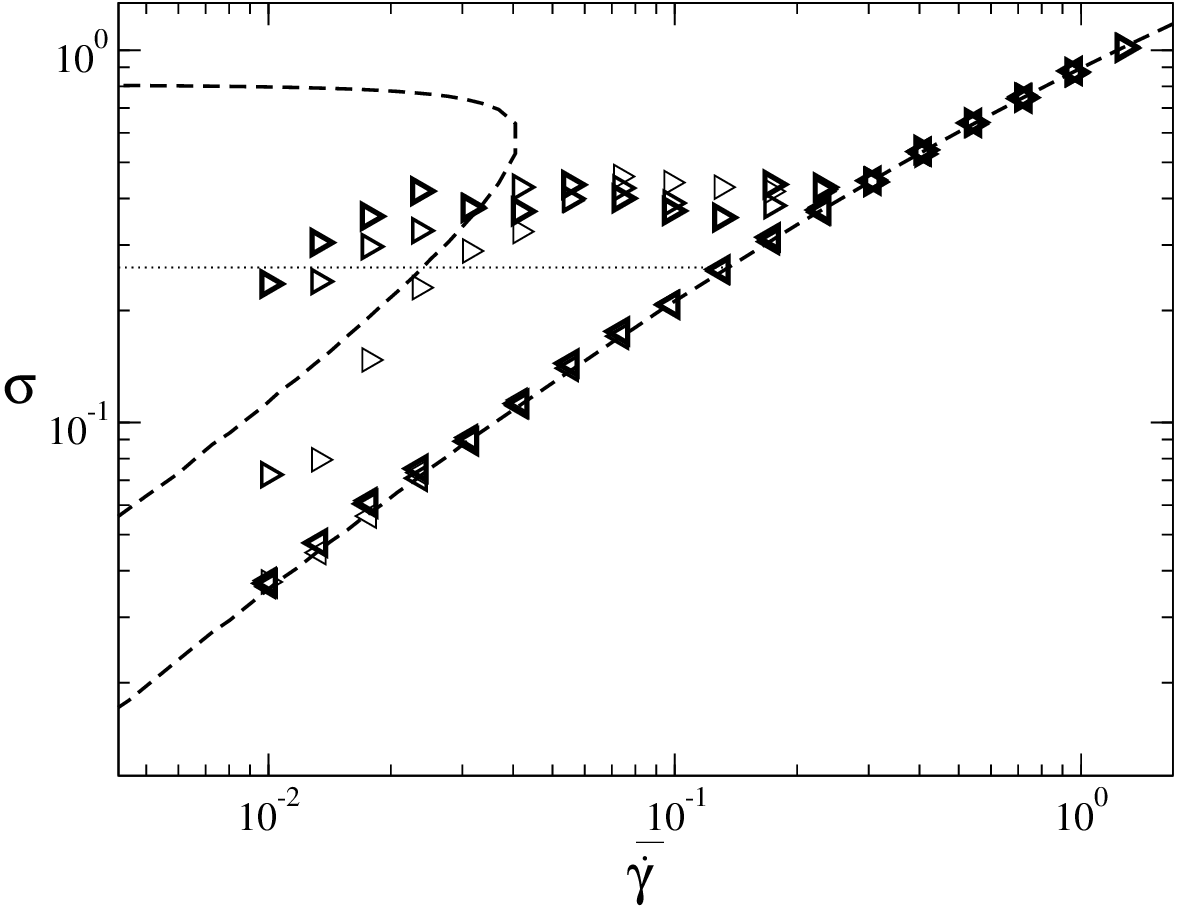}}
\caption{Dashed lines: constitutive curve $\sigma(\dot\gamma)$ for Model 2 with $x_0=0.15,\tilde a=3.75$. Symbols: WTMC data ($n=100, m = 1000, \lambda = 0.5\Delta$) for up/down strain-rate sweeps (left-pointing and right-pointing triangles respectively); $t_r = 200,400,800$ (thin, medium, bold symbols), initialized in a homogeneously aged state of $t_w = 10^4$.
Dotted line, as guide to the eye, the quasi-steady stress attained at long times in shear startup for $\overline{\dot\gamma} \le 0.1$ in the shear banding regime.} 
\label{fifth}
\end{figure}

In what follows we explore two model variants, distinguished
according to different choices for the dependence of $S$ upon
$P(y)$.  Model 1 has
\begin{equation} {\cal S}(y)  = a\langle\ell^2/\tau\rangle_{P(y)}
\label{one}
\end{equation}
where $\tau = \exp[(E-k\ell^2/2)/x]$ is the (dimensionless) trap lifetime. 
In this model the noise is pumped by dissipation of elastic energy. Model 2 instead has ${\cal S}(y) = \tilde a\langle
1/\tau\rangle_{P(y)}$ such that all jumps contribute equally to noise
regardless of the local strain released. As discussed below, the underlying constitutive curve of both models has a
vertical branch at $\dot\gamma=0$, terminating in a yield stress; and
a fluid branch that persists to stresses below this yield value
(Figs.~\ref{first} and~\ref{fifth}). Both models are thereby capable
of capturing a coexistence of unsheared and fluid bands at a common
shear stress. Indeed, many of the flow phenomena that are for
concreteness discussed below in the context of model 1 can in
principle also arise in model 2. Beyond these common phenomena,  model 2 additionally
allows the fluid branch to persist right down to the origin, allowing it to capture, for instance, the large hysteresis loops seen during shear rate
sweeps in glassy star polymers (Fig.~\ref{fifth} below).

In the second equality of (\ref{xdynamics}) we have for simplicity set $\tau_x\to 0$, so that local jump rates adapt rapidly to changes in nearby activity levels. In the same spirit, we neglect fluid inertia and therefore impose force balance, which, for planar Couette flow, requires uniformity of the shear stress: 
$\sigma(y) = k\langle\ell\rangle_{P(y)} = \sigma$, a constant.

The diffusive term in (\ref{xdynamics}) represents the nonlocal effect of jumps. More generally one might replace ${\cal S}$ in (\ref{xdynamics}) by $\int dy'dt'{\cal S}(y',t'){\cal G}(y-y',t-t')$ where $\cal G$ is a kernel. We assume that $\cal G$ is effectively of short range: most of the noise comes from nearby jumps. (Note that, despite the long-range nature of elastic interactions, the stress field caused by a randomly signed sum of plastic strains distributed through space is likewise dominated by local contributions \cite{ajdari}.) If one further assumes that a smoothly varying mean activity level governs the rates of individual stochastic events, a gradient expansion in activity is thereby justified. To obtain (\ref{xdynamics}) as written, we take the leading order correction within this expansion, ${\cal S} \to {\cal S} + \lambda^2\partial^2{\cal S}/\partial y^2$, and iterate once to set $\partial^2{\cal S}/\partial y^2\to \partial^2x/\partial y^2$. The latter replacement should be harmless, and greatly simplifies the numerical analysis.

To solve the above systems numerically we discretize into $i=1...n$ streamlines equally spaced in the $y$ direction with a spacing $\Delta$. We take on each streamline a separate ensemble of $j=1...m$ SGR elements with barrier heights $E_{ij}$. The stress on streamline $i$ is $\sigma_i = (k/m)\sum_j \ell_{ij}$. Periodic boundary conditions are used, and force balance is imposed as follows. Suppose a jump occurs at element $ij$ when its local strain is $\ell = l$. By updating all elements on the same streamline as $\ell\to \ell + l/m$, force balance is maintained, but with a stress level that is unchanged by the jump. Further updating all elements throughout the system as $\ell \to \ell -l/mn$ restores the global stress to the proper reduced level. This algorithm can also be thought of as the $\eta_s \to 0$ limit of one in which a small Newtonian viscosity $\eta_s$ is introduced alongside the elastic stress of SGR elements, and local strain rates are set to maintain $\sigma(y)+\eta_s\dot\gamma = \sigma$ at all times. The SGR sector of the numerics is handled by a waiting-time Monte Carlo (WTMC) algorithm \cite{wtmc} that chooses stochastically both the element and timing of the next yield event. After each such event, force balance is applied; every $N$ WTMC steps, (\ref{xdynamics}) is then evolved to steady state (we found $N\simeq 10$ sufficient for accuracy, and more efficient than $N=1$).

\begin{figure}
\centerline {\includegraphics[width=2.6in]
{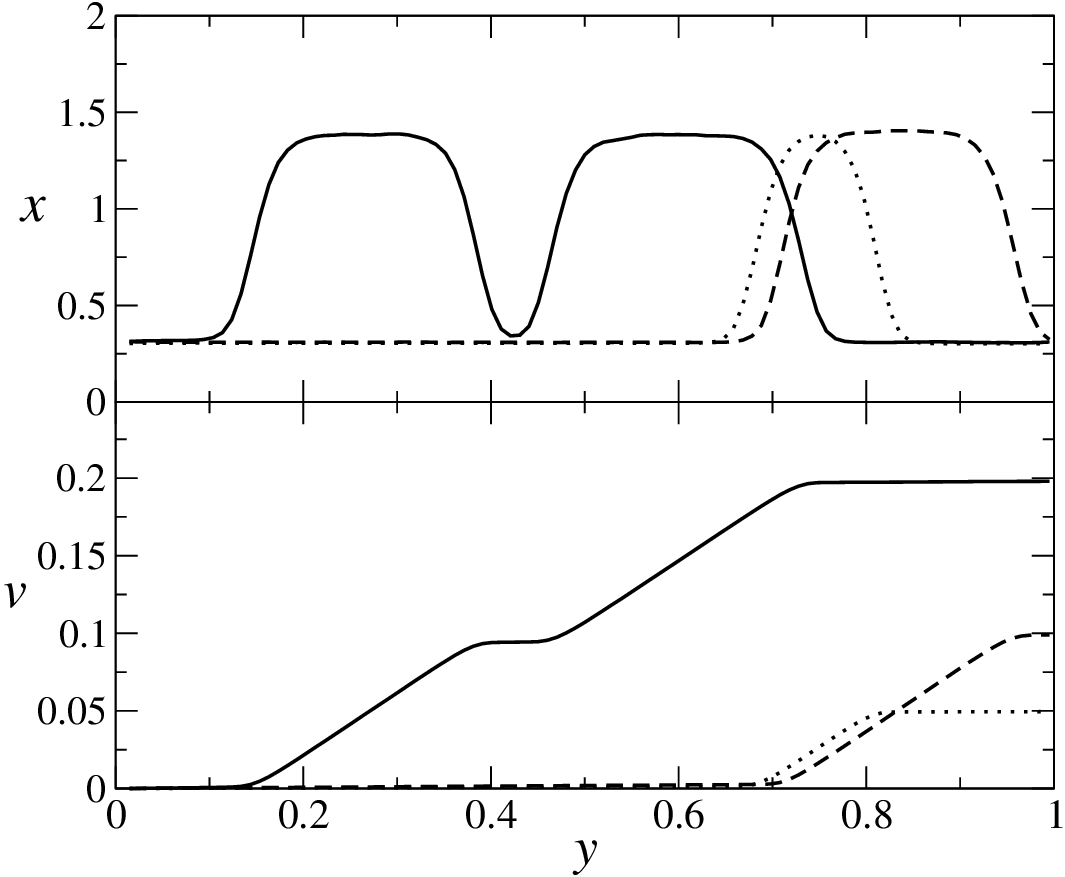}}
\caption{Profiles of noise temperature $x(y)$ and velocity $v(y)$ for $x_0=0.3$ and $a=2.0$ with imposed mean shear rate $\overline{\dot\gamma} = 0.05, 0.1, 0.2$ (dotted, dashed, solid). Data is time averaged over the second half of a run with total strain 100. } 
\label{third}
\end{figure}

We now discuss our results for Model 1, letting $\Gamma_0 = k = 1$  define time, strain and stress units. First, note that in a homogeneous steady state (\ref{xdynamics}) is equivalent to
\begin{equation}
x = x_0 + 2a\sigma(x,\dot\gamma)\dot\gamma \label{homflow}
\end{equation}
which allows us to generate constitutive curves $\sigma(\dot\gamma)$ from the SGR constitutive equation \cite{SGR} without recourse to WTMC. The result is shown in Fig.\ref{first} for the case $x_0=0.3$ and $a = 2.0$. Also shown is our WTMC data for the actual flow behavior in planar Couette flow. This is quasi-stationary data, taken during a strain window $150.0\le\gamma\le 200.0$ following startup of steady shear.
As in any strain-controlled experiment, only the {\em mean} shear rate $\overline{\dot\gamma}$ is imposed: the system is free to choose a banded state, and indeed it does so for a wide window of strain rates (Figs.\ref{first},\ref{third}). All these data were generated with $n=100,m=1000$ and $\lambda/\Delta = 0.5$ or $1.0$ (sufficient to resolve the interface). For each run the sample is `fresh', quenched from a state with $P(E,\ell) \sim \rho(E)\delta(\ell)$ at $t=t_w\simeq 1$; waiting for a time $t_w\gg 1$ at $x=x_0$ before startup of shear yields a larger stress overshoot but the same plateau stress. Note that the multiple bands seen at $\overline{\dot\gamma}=0.2$  in Fig.~\ref{third} can only arise in a planar shear flow: they would presumably be eliminated by any small curvature in the flow geometry, as is almost always present experimentally.

The results in Fig.\ref{first} are consistent with the following picture. The homogeneous flow curve has a yield stress $\sigma_Y = \sigma_Y(x_0)$ inherited from the simple SGR model. Unlike simple SGR, however, $\sigma(\dot\gamma>0)$ is thereafter a {\em decreasing} function, before re-stabilizing at higher shear rates. 
(In fact the initial slope of the curve is positive, but for the chosen parameters $\sigma(\dot\gamma)$ has a maximum at $\sigma-\sigma_Y \simeq 2 \times 10^{-4}$, invisible in Fig.\ref{first}, and falls back below $\sigma_Y$ for $\dot\gamma\ge 10^{-3}$.)
This creates the standard conditions for shear banding but, unusually, the more viscous band has $\dot\gamma =0$ \cite{coussot2}. This band is thus effectively solid, with local strain rate that is close to zero (Fig.\ref{third}), and slowly decreasing with time. The latter represents an aging effect: this band has a low noise temperature $x \simeq x_0= 0.3$ such that $\sigma_Y(x)>\sigma$, which is the condition for nonergodicity to arise within SGR. This aging can be confirmed directly by study of the correlator $C(t,t_w)$ (the fraction of unhopped particles) within the low shear band. This shows simple aging behavior, as does the standard SGR model at the given $x=x_0$ \cite{SGRage}. In contrast, the high-shear band has high activity, self-consistently maintaining it in an ergodic state of high $x$ and low viscosity. 

\begin{figure}
\centerline {\includegraphics[width=2.8in]
{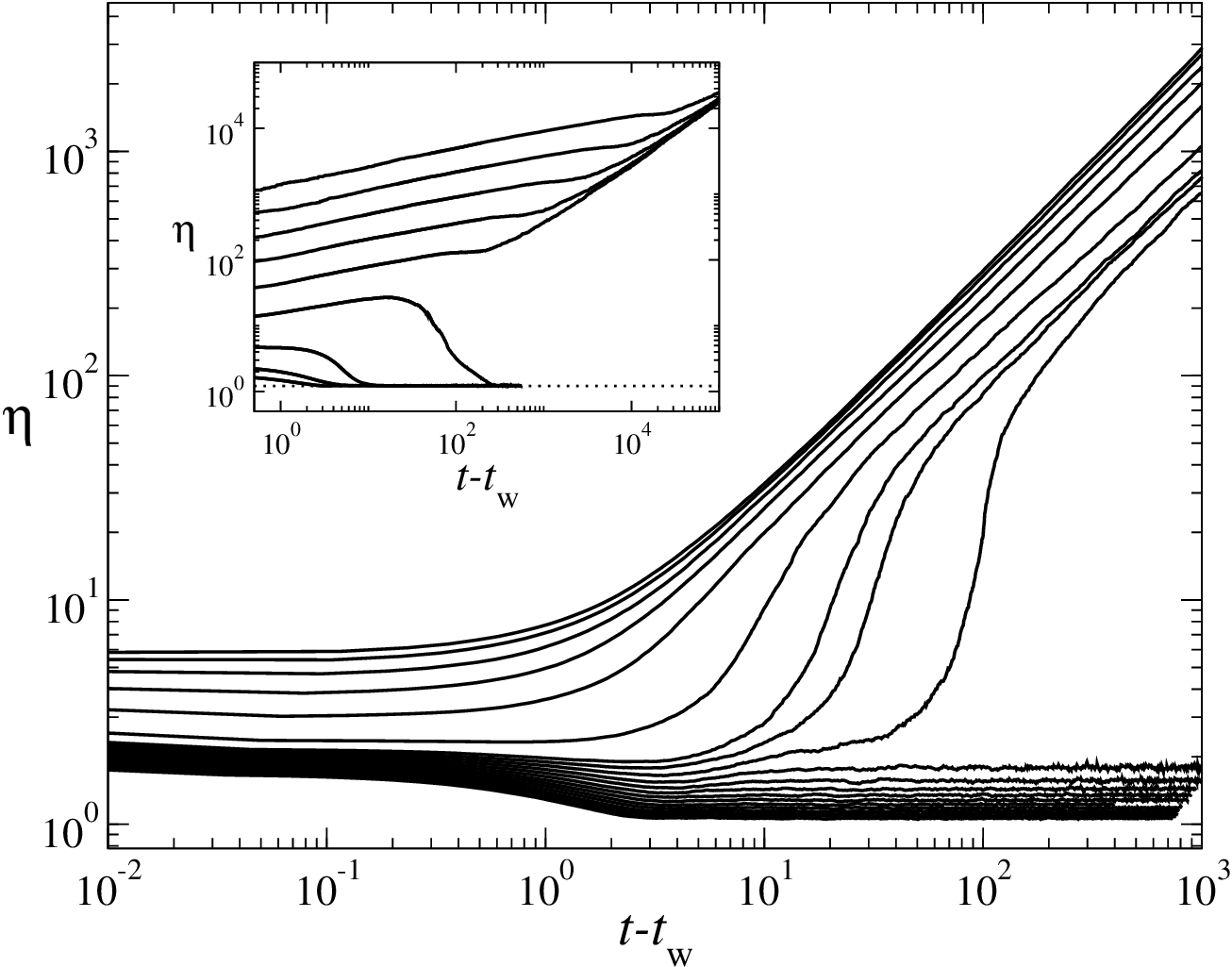}}
\caption{Viscosity bifurcation for $x_0 = 0.3, a = 2.0, n=100, m = 1000$. Main figure: age before shear $t_w = 1$; stress values (top to bottom at right) $\sigma = 0.1,0.2,0.3,0.4, 0.5, 0.6, 0.64, 0.65, 0.66, 0,67, 0.68, 0.69, 0.70,$ $ 0.71, 0.72, 0.73, 0.74, 0.75, 0.76, 0.77, 0.8$. Inset: stress $\sigma = 0.72$, age before shear $\log_{10}t_w = 0,0.5,1,1.5,2,2.5,3,3.5,4$. Dotted line shows the asymptote calculated semi-analytically.} 
\label{fourth}
\end{figure}

 This scenario is supported by simulations on molecular models
\cite{coussot2}, and is consistent with experiments on
viscosity-bifurcating materials \cite{coussot,bonn}. Indeed, we show
in Fig.\ref{fourth} a plot of viscosity $\eta = \sigma/\dot\gamma$
against time $t$ following a step stress on fresh ($t_w=1$)
samples. As expected, without fine-tuning of $\sigma$, we find flows
that are homogeneous. There is a clear bifurcation between
$\sigma<\sigma_c$, for which $\eta(t)$ is large and increases without
limit with $t$, and $\sigma> \sigma_c$ for which $\eta(t)$ falls onto
a plateau. Close to $\sigma_c$ the time to nucleate the non-flowing
branch is long for large systems; but assuming this remains finite our
best numerical estimate for fresh samples is $\sigma_{c,f} = 0.675\pm
0.005$. Intriguingly, this is very close to the shear-banding stress
plateau found in Fig.\ref{first} and well {\em below} the dynamic
yield stress $\sigma_Y(x_0)$ for homogeneous flow. For fresh samples,
the strain-rate overshoot following step stress at
$\sigma_{c,f}<\sigma<\sigma_Y$ carries the system onto the fluid
branch, where it remains.
 
The inset to Fig.\ref{fourth} shows another intriguing facet of the viscosity bifurcation: the critical stress is itself an increasing function of $t_w$ (see also Fig.\ref{first}).  Thus, if subjected to step stress $\sigma_{c,f}<\sigma<\sigma_Y(x_0)$, a series of samples of different ages will show a bifurcation between flow (small $t_w$) and arrest (large $t_w$). In an SGR context this is natural: in an aged sample, elements have fallen into deeper traps, so that the strain-rate overshoot is more modest, and the transient yield events caused by the step stress do not receive enough feedback to cause runaway to a fluid state. Although not emphasized in \cite{coussot}, a similar $t_w$ bifurcation holds for the phenomenological model reported there, in which an inverse fluidity or `jamming parameter', $\lambda$, evolves with time. Our SGR results broadly support the idea \cite{coussot} that this evolution represents aging.

We next turn to Model 2, focussing on a specific regime that may be relevant to experiments on star polymers \cite{stars} which showed, under relatively rapid strain-rate sweep (residence time $t_r = 10$s per observation point), an apparently conventional, monotonic flow curve. However, for slower sweeps ($t_r =10^4$s) a much larger, almost constant, stress was measured at small $\overline{\dot\gamma}$. Shear banding in this region was confirmed by NMR velocimetry, with $\dot\gamma\simeq 0$ in the slow band \cite{stars2}. 
A strong hysteresis was also seen, the less viscous branch persisting to much lower strain rates on the downward sweep \cite{stars}.  

Fig.~\ref{fifth} shows data for Model 2 with $x_0 = 0.15, \tilde a =
3.75$. The right-pointing triangles show the stress response to a slow
upward strain rate sweep for a sample that was in a homogeneously aged
state of $t_w=10^4$ before shear. A stress plateau is clearly seen for
shear rates $\dot\gamma < 0.3$. In this regime, the system forms
coexisting glassy and flowing shear bands.  At the lowest applied
shear rates, the stress does not have time fully to attain the plateau
value before the strain rate is swept on to a higher value. This
accounts for the reduced stresses at the far left hand edge of the
plot. As would be anticipated, this reduction is pushed to smaller
shear rates for slower ramp rates. For shear rates $\dot\gamma>0.3$
the system flows homogeneously on the fluid branch of the constitutive
curve.

A remarkable feature of Model 2, not seen in Model 1, is that the
constitutive curve $\sigma(\dot\gamma)$ remains multivalued down to
$\dot\gamma = 0$ where a quiescent glass at $x=x_0$ and a fluid phase
at $x_1 = x_0 + \tilde a \Gamma(x_1)/\Gamma_0$ both exist
\cite{footnote}. If a system is prepared on the fluid branch at high
$\dot\gamma$, then barring macroscopic nucleation events (which do not
occur at measurable rate in our numerics), the shear rate can be
ramped down to zero while maintaining homogeneous fluidity. The same
applies to the downward part of an upward then downward shear rate
ramp, as shown by the left-pointing triangles in
Fig.~\ref{fifth}. Apart from the absence of nucleation, which we
assume would cause a sweep-rate-dependent escape from the fluid branch
on down-ramping the strain-rate, the WTMC data in Fig.\ref{fifth}
intriguingly resembles that reported experimentally in \cite{stars}.

Note more generally that nucleation events are
hard to capture in numerics (see e.g.~\cite{FFS}) and their absence
from Model 1 might give, for instance, small shifts in the
$t_w$-dependent $\sigma_c$ found there. However, the qualitative
features of both models appear robust to such effects. Numerical
difficulties also impede detailed study of the limit $x_0\to 0$, which
might be the realistic limit when true thermal noise is negligible
\cite{SGR}. While qualitatively similar flow curves should arise, this
limit of the standard SGR model is semi-deterministic (the element of
lowest barrier height is always next to yield -- reminiscent of
extremal models of self-organized criticality) so that the treatment
of yield-induced noise becomes more delicate. 
 
 In conclusion, we
have presented an extended SGR model for soft glasses under flow, in
which the noise temperature $x$ varies to reflect the dependence of
local jump rates on yield events elsewhere within the sytem. Two
versions of the model offer a connection to the physics of viscosity
bifurcations (Model 1) and star polymer colloids (Model 2), neither of
which could previously be accounted for within a simple (uniform $x$)
SGR framework. So far, our numerical work has addressed only the case of periodic boundary conditions. Very recent experiments however emphasise the perturbing influence of boundaries \cite{recent}, which might or might not be linked to their effects on the noise temperature; we hope to explore this in future work. Meanwhile, we believe that mesoscopic models, such as those
developed here, continue to offer a useful compromise between first
principles \cite{catesfuchs} and fully phenomenological approaches
\cite{picard,coussot,stars} to the rheology of nonergodic soft
matter.
 
 SMF thanks EPSRC EP/E5336X/1 for funding; MEC holds a
Royal 
 Society Research Professorship. SMF would like to thank Paul Callaghan for his hospitality during a research visit to the Victoria University of Wellington, where part of this work was carried out.


\begin{thebibliography}{99}
\bibitem{footnote} An upper fluid branch visible in Fig.\ref{fifth} is presumably unstable; slight parameter variation allows the two fluid branches to merge at small $\dot\gamma$, which is perhaps more physical, without otherwise changing the scenario reported here.
\end{thebibliography}
\end{document}